\documentclass[3p,times,procedia]{elsarticle}
\usepackage{nupha_ecrc}

\volume{00}

\firstpage{1}

\journalname{Nuclear Physics A}

\jid{nupha}

\jnltitlelogo{Nuclear Physics A}

\usepackage{amssymb}

\begin{document}

\begin{frontmatter}

\dochead{XXVIIIth International Conference on Ultrarelativistic Nucleus-Nucleus Collisions\\ (Quark Matter 2019)}

\title{Electromagnetic \& Weak Probes: Experimental Overview}

\author{Frank Geurts}

\address{Department of Physics \& Astronomy, Rice University, Houston TX 77005, USA}

\begin{abstract}
Electromagnetic and electroweak probes are the most versatile probes in the study of heavy-ion collisions. Produced at 
every stage in the evolution of QCD matter, its messengers are practically inert to the strongly interacting medium 
they travel through. In this contribution, I will discuss a selection of new results from experiments at the LHC, RHIC,
and SIS facilities, spanning almost four orders of magnitude in beam energy. I will conclude with a brief overview of 
the experimental landscape in the near future.
\end{abstract}


\end{frontmatter}

\section{Introduction}
Colorless probes are the ideal carriers of information in a strongly interacting medium. While not
completely immune to interactions with hadrons, or quarks for that matter, the typical mean free 
path of leptons and photons is long enough to allow these particles to escape and make it to our 
detectors. Moreover, the reach into the TeV scale has made electroweak bosons the latest
asset in the experimentalist's toolbox, helping to provide constraints on nuclear parton
distributions functions in p$+$A and A$+$A collisions. Thus, electromagnetic and weak probes can provide a
complete picture with access to initial conditions and the subsequent evolution of the system; ranging
from hard process, such as the $W^\pm$, $Z$ production and prompt photons, to soft
process that include thermal photons and dileptons originating from leptonic decays and virtual photons. 

The present experimental landscape of electroweak probes involves several facilities and many more experiments
spanning almost four orders of magnitude in center-of-mass energies. The versatility of these facilities
provided for electroweak and dilepton data from a wide range  of collision systems, ranging from heavy-ion 
collisions such as Pb$+$Pb and Au$+$Au, to smaller systems such as In$+$In and Cu$+$Cu, p$+$Pb, and pp.
Detector improvements furthermore opened up access to a range of collision geometries from ultra-peripheral
collisions (UPC), to nuclear overlaps in peripheral hadronic collisions, and high-multiplicity pp collisions.
It is a testament of a very active field to see at this conference so many results as well as plans for the 
future presented. In what follows only a small selection of several exciting results can be discussed.

\section{Electroweak Bosons}
\label{sect:ewbosons}
Electroweak bosons are created in high-momentum processes that take place in the initial stages of a
collisions. This makes these bosons excellent probes of the parton distribution functions (PDFs) inside a 
nucleon. Moreover, the electroweak boson's fast decay into leptons allows the information of the partonic 
structure of the initial state to be carried without being affected by a strongly interacting medium in
the case of heavy ion collisions. The production of W bosons, primarily through 
$\mathrm{q}\bar\mathrm{q}$ annihilation, provides access to the distribution of the light (anti-) quarks. 
Proton PDFs, based on global fit analyses, now include data from recent LHC results. In a bound nucleus, 
these distribution functions are observed to be modified. The nuclear PDFs can be expressed in terms of 
$f_i^{p/A}(x,Q^2)=R^A_i(x,Q^2)f_i^p(x,Q^2)$, where the nuclear modification $R^A_i(x,Q^2)$ shows at small 
Bjorken $x$ the effect of shadowing, at intermediate $x$ that of anti-shadowing and the EMC effect, at 
high $x$ followed by Fermi motion \cite{Armesto:2006ph}. 
\begin{figure}[t]
    \centering
    \includegraphics[width=0.46\textwidth]{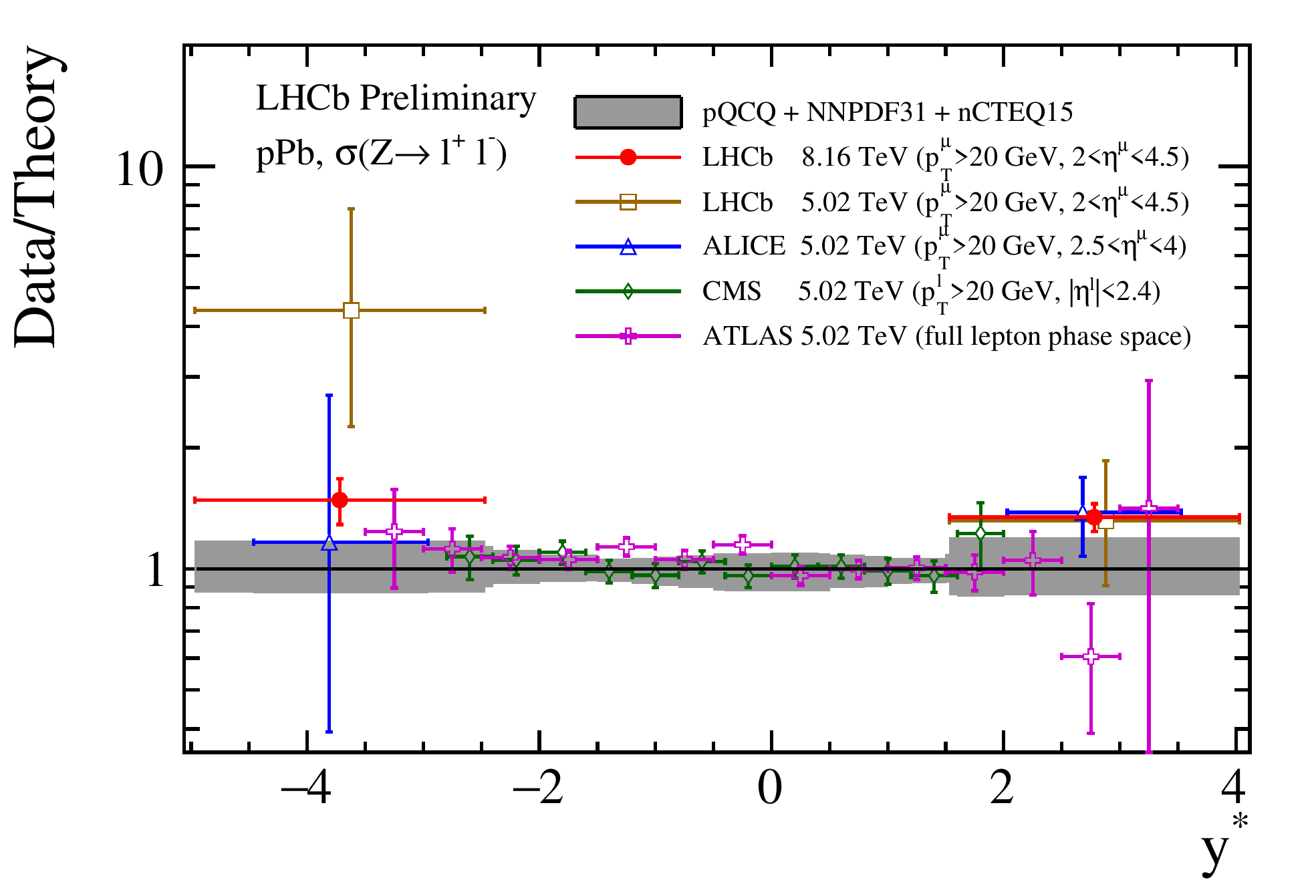}
    \includegraphics[width=0.46\textwidth]{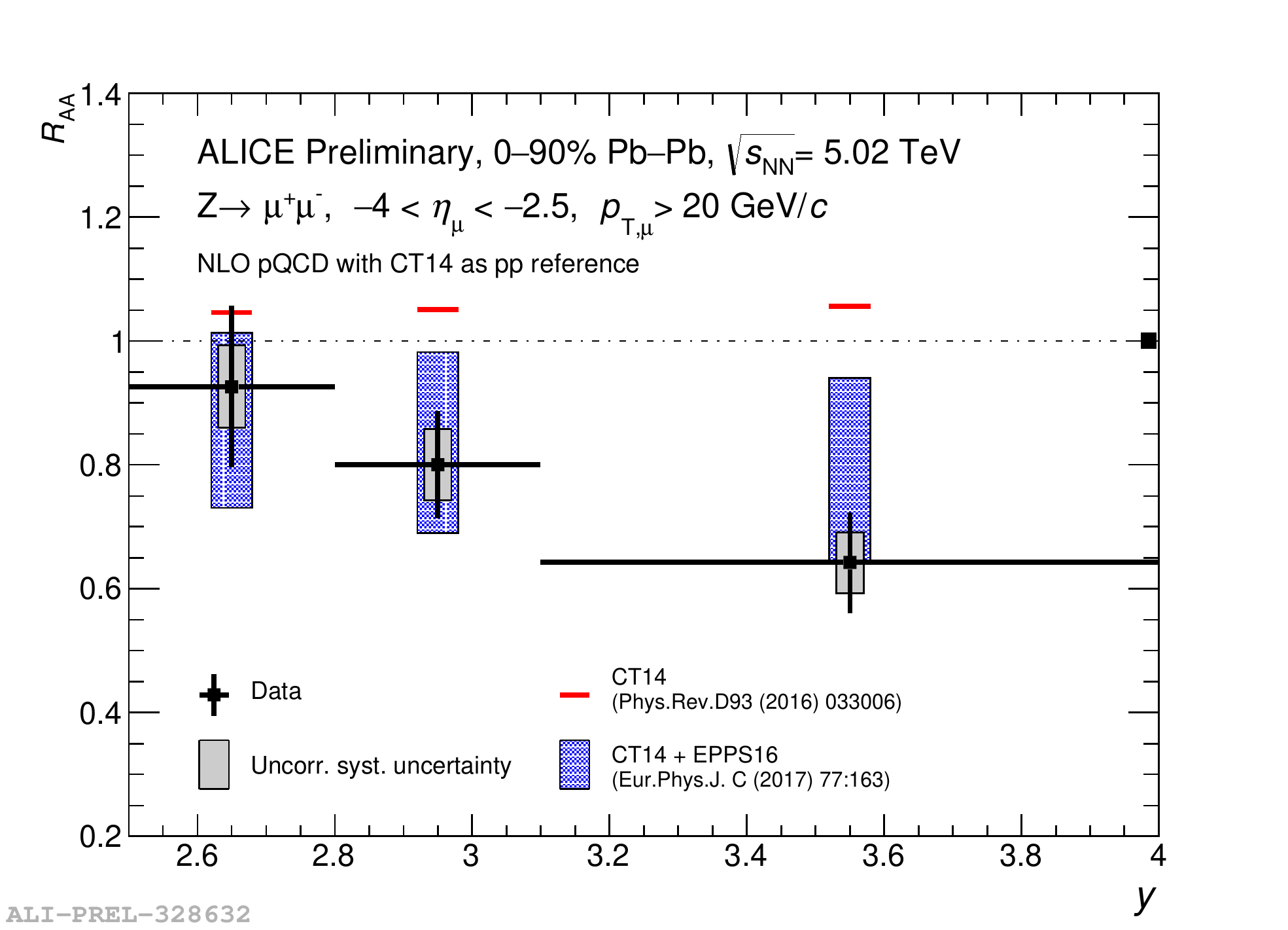}
    \caption{Electroweak boson production in p$+$Pb and Pb$+$Pb collisions. Left panel: Comparison of results on Z-boson
    production in p$+$Pb collisions at $\sqrt{s_{NN}}=8.16$~TeV measured by LHCb, compared with previous results from ATLAS,
    CMS, ALICE, and LHCb at $\sqrt{s_\mathrm{NN}}=5.02$~TeV \cite{qm2019:LHCb:Li}. Right panel: Z-boson production in Pb$+$Pb at $\sqrt{s_\mathrm{NN}}=5.02$~TeV as a function of dimuon $y$, compared with CT14 PDF and EPPS16 nPDF \cite{qm2019:ALICE:Weber}.}
    \label{fig:LHC_Zbosons}
\end{figure}

At this conference, the CMS and ATLAS collaborations reported on its recently published results of $W^\pm$ 
production from p+Pb collisions at $\sqrt{s}=8.16$~TeV \cite{Sirunyan:2019dox,Aad:2019sfe}. These results 
not only demonstrate that nPDFs are clearly favored over the CT14 PDF but with experimental uncertainties 
smaller than those of the models it is possible  to differentiate between EPPS16 and CTEQ15. New results 
from Z boson measurements in p-Pb at $\sqrt{s}=8.16$~TeV and Pb-Pb at $\sqrt{s}=5.02$~TeV by the LHCb and 
ALICE collaborations, respectively, are shown in Fig.\ \ref{fig:LHC_Zbosons} and were shown to be 
compatible with CTEQ15 and EPPS16 predictions.

\section{Direct Photons}
\label{sect:directphotons}
\begin{figure}[t]
    \centering
    \includegraphics[width=.59\textwidth]{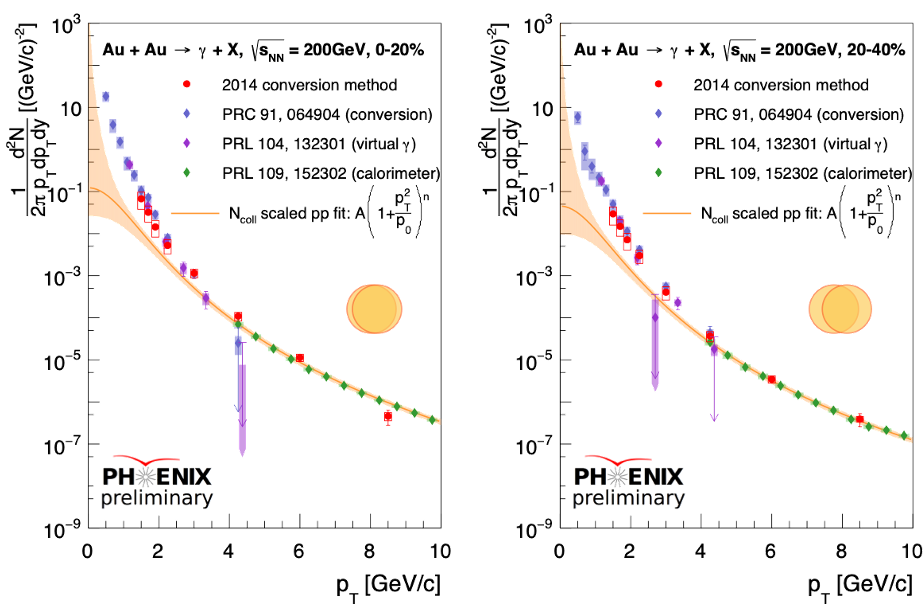}
    \includegraphics[width=.4\textwidth]{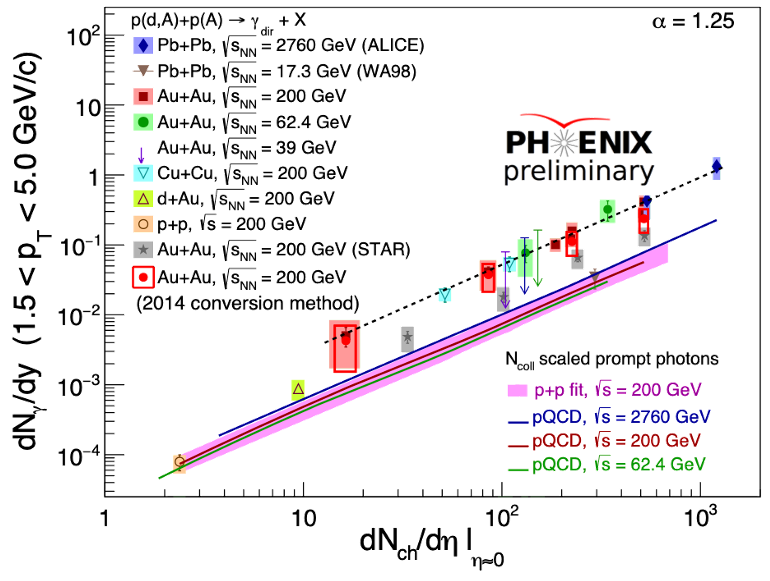}
    \caption{Direct photon yields in Au$+$Au collisions at $\sqrt{s_\mathrm{NN}}=200$~GeV \cite{qm2019:PHENIX:Fan}. 
    Left panel: for 0-20\% central collisions. Middle panel: for 20-40\% semi-peripheral collisions. Right panel: 
    universal scaling of low-$p_\mathrm{T}$ direct photon yields $dN_\gamma/dy$ with respect to the number of charged 
    particles at midrapidity $dN_\mathrm{ch}/d\eta|_{\eta=0}$.}
    \label{fig:phenixdirectgamma}
\end{figure}

Photons are produced throughout the entire space-time evolution of a strongly interacting system.
Its sources include prompt hard parton scattering, thermal radiation, and jets - collectively referred to as direct 
photons \cite{David:2019wpt}. In addition, photons can originate from the electromagnetic decay of final state 
hadrons. The major experimental challenge is to disentangle these contributions. Electromagnetic decays from final 
state hadrons by far provide the largest contribution and as such form a substantial backgrounds to the
measurement of direct photons. The PHENIX collaboration reported on new measurements of $R_\gamma(p_\mathrm{T})$ in
Au$+$Au at $\sqrt{s_\mathrm{NN}}=$200~GeV. These new measurements, shown with red symbols in the two left panels of 
Fig.~\ref{fig:phenixdirectgamma} for different centralities use the external conversion  method and are based on 
the large RHIC Run-14 data sample.
The results show a clear enhancement in the direct photon yields for $p_\mathrm{T} \le 3$~GeV/$c$, which continues to 
persist in the semi-peripheral data (middle panel). At high momenta, the results show consistency with
$N_\mathrm{coll}$-scaled p$+$p results. In another important consistency check, PHENIX's new results show good
agreement with previously published results that based on different data sets \cite{Adare:2014fwh}, and/or different methods
such as the virtual-$\gamma$ \cite{Adare:2008ab} and the calorimeter methods \cite{Adare:2011zr}. In the right panel of
Fig.~\ref{fig:phenixdirectgamma}, the invariant yield of photons is plotted as a function of the charged hadron 
multiplicity $dN_\mathrm{ch}/d\eta$ at midrapidity. The new data from PHENIX are in line with the recently observed 
scaling \cite{Adare:2018wgc}. In the same figure, data from the STAR experiment is added. While the scaling appears to be
similar, the rates are systematically lower. It is expected that data from the ongoing STAR Beam Energy Scan (BES) Phase-2
program, see Sect.~\ref{sect:future}, will be able to add several new points at the lower charged hadron multiplicities
using a similar conversion technique \cite{STAR:2016use}.

\section{Dileptons}
\label{sect:dileptons}
At this conference, a wealth of new experimental dilepton data has been being presented comprising more than three
orders of magnitude in $\sqrt{s_\mathrm{NN}}$. Dilepton invariant-mass spectra bring a plethora of physics channels
from different stages of the evolution of the medium that can be "tuned in" by selecting the relevant mass window 
and thus not only include the leptonic decay channels of various light, strange, and charm mesons but also allow
for the measurement of virtual direct photons from similar sources as mentioned previously in Sect.~\ref{sect:directphotons}. 
The HADES collaboration recently published \cite{Adamczewski-Musch:2019byl} its results from measurements of dielectron 
production in Au$+$Au collisions at $\sqrt{s_{NN}}=2.42$~GeV . Its findings confirm at this energy the strong in-medium 
modification of the $\rho$ meson, first reported at SPS energies by the NA60 collaboration \cite{Arnaldi:2006jq}.
\begin{figure}[h]
    \centering
    \includegraphics[width=0.98\textwidth]{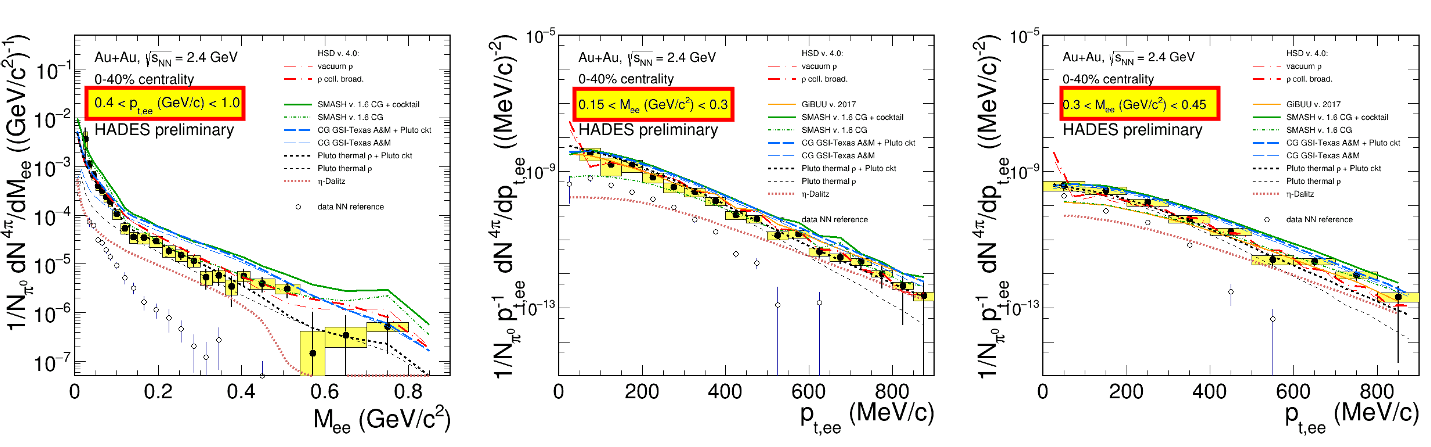}
    \caption{Thermal dielectron measurements from Au$+$Au collisions at $\sqrt{s_\mathrm{NN}}=2.42$~GeV by the 
    HADES collaboration \cite{qm2019:HADES:Harabaz}. Left panel: invariant-mass spectra for low-$p_\mathrm{T}$ 
    dielectrons. Middle and right panels: transverse momentum distributions for $0.15 < M_{ee} < 0.3$~GeV/$c^2$ 
    and $0.3 < M_{ee} < 0.45$~GeV/$c^2$, respectively.}
    \label{fig:HADESee}
\end{figure}{}
After careful removal of the hadronic contributions to the invariant mass spectrum the HADES collaboration 
extracted, based on a black-body spectral function fit, the average temperature of the radiating fireball of $71.8\pm2.1$~MeV.
At this conference, the collaboration reported on its multi-differential measurements of the dielectron invariant-mass 
and $p_\mathrm{T}$ spectra \cite{qm2019:HADES:Harabaz}. In Fig.~\ref{fig:HADESee}, a comparison of the invariant-mass 
yield (left panel) and momentum spectra in two mass windows (middle and right panels) are compared with several model 
descriptions and shows to have sufficient sensitivity to the details of the model descriptions.

Dilepton-based measurements of the azimuthal anisotropy $v_2$ as a function of $p_\mathrm{T}$ in different invariant mass regions
have been long been proposed as an alternative way to study medium at the different stages \cite{Chatterjee:2007xk}.
However, measuring the dielectron $v_2$ is a statistics-hungry challenge, see e.g.~\cite{Adamczyk:2014lpa}. Here, HADES
reported on its first findings based on 2.6 billion events, and found a consistent comparison of its preliminary results in 
the $\pi^0$ Dalitz mass range the $v_2$ of charged pions \cite{qm2019:HADES:Harabaz}.

\begin{figure}[h!]
    \centering
    \includegraphics[width=0.35\textwidth]{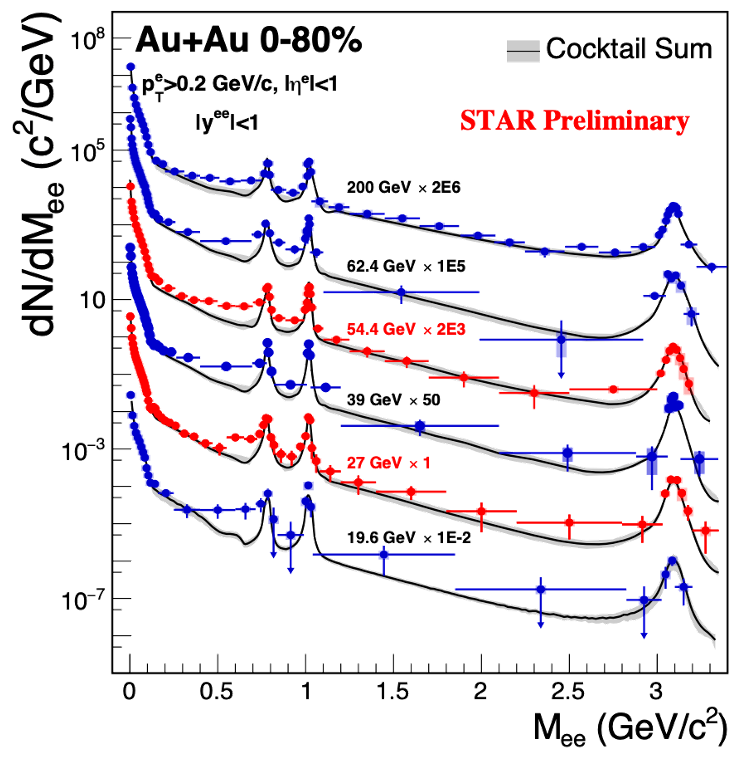}
    \includegraphics[width=0.3\textwidth]{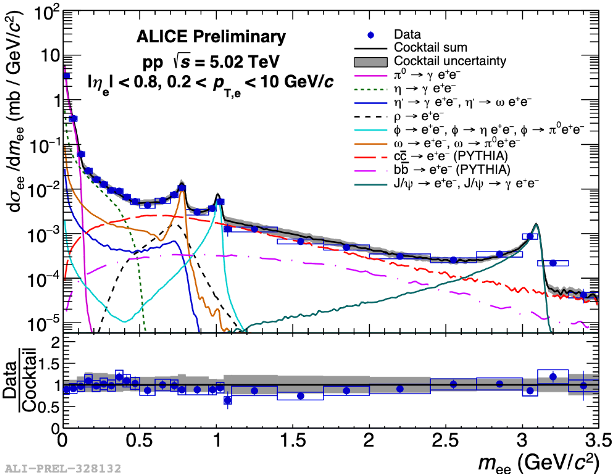}
    \includegraphics[width=0.3\textwidth]{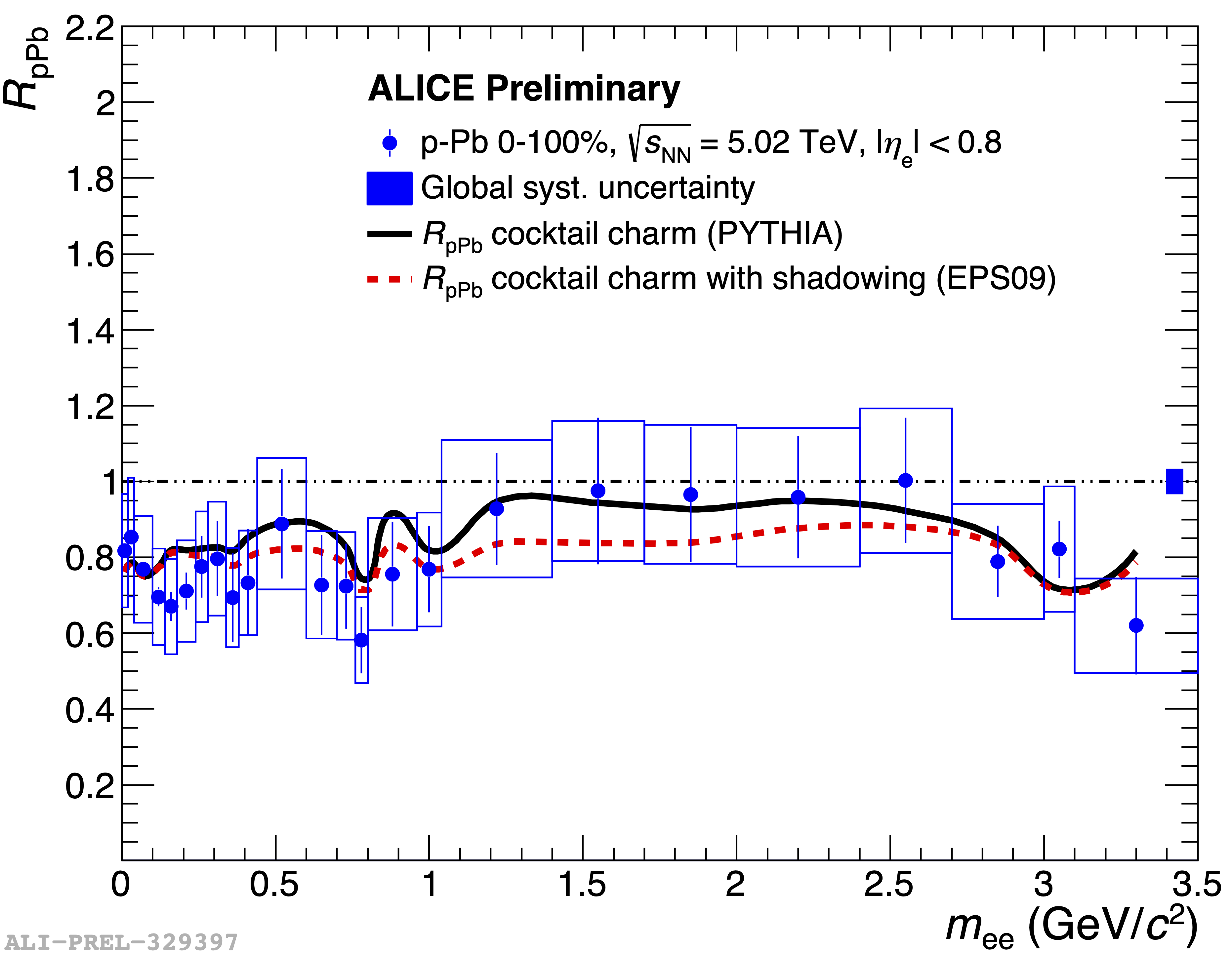}
    \caption{Dielectron invariant mass spectra in Au$+$Au and Pb$+$Pb collisions from STAR and ALICE, respectively. 
    Left panel: high-statistics measurements at $\sqrt{s_\mathrm{NN}}=27$ and 54.4~GeV (red symbols) by STAR \cite{qm2019:STAR:Seck}.
    Middle and right panels: ALICE results for p$+$p and $R_\mathrm{pPb}$ in collisions at $\sqrt{s_\mathrm{NN}}=5.02$~TeV, respectively \cite{qm2019:ALICE:Scheid}.}
    \label{fig:dilepton}
\end{figure}
The STAR collaboration reported its new dielectron results from three high-statistics data sets \cite{qm2019:STAR:Seck}.
In the left panel of Fig.~\ref{fig:dilepton} the dielectron invariant mass for the $\sqrt{s_\mathrm{NN}}=27$~GeV and 54.4~GeV
energies is overlaid with STAR's results from other collision energies, including the first RHIC BES \cite{Adam:2018qev}.
A ten-fold increase in event statistics compared to the BES data is expected to better constrain the cocktail by 
direct measurements of the $\omega$ and $\phi$ mesons, and allow for virtual direct photon measurements. The uncertainties
in these new results are a considered good indicators of the expected precision for the BES Phase-2 energies between
$\sqrt{s_\mathrm{NN}}=7.7$ and 19.6~GeV. 

In the middle panel of Fig.~\ref{fig:dilepton}, new dielectron invariant-mass results from the ALICE 
collaboration are shown for p$+$p collisions at $\sqrt{s}=5.02$~TeV. The vacuum baseline in the
p$+$p data is found to be well described by the expectations from the hadronic cocktail. The distinct shape of the charm
and beauty contributions in the intermediate mass range ($1.1\le M_\mathrm{ee} \le 2.7~$GeV/$c^2$) is used to extract the charm
and beauty cross sections which is found to be consistent with independent heavy-flavor measurements
\cite{qm2019:ALICE:Scheid}. In the right panel of Fig.~\ref{fig:dilepton}, the ALICE collaboration used the p$+$Pb 
invariant mass spectra to verify initial state nuclear modification, $R_\mathrm{pPb}=\frac{1}{\langle
N_\mathrm{coll}\rangle}\frac{dN/dM_\mathrm{ee}|_\mathrm{pPb}}{dN/dM_\mathrm{ee}|_\mathrm{pp}}$ at $\sqrt{s_\mathrm{NN}}=5.02$~TeV.
In the intermediate mass range, the results do not show significant modifications in agreement with previous D-meson
measurements from the ALICE collaboration. However, in the low mass range ($M_\mathrm{ee} \le 1$~GeV/$c^2$ a deviation
from unity is observed. This deviation is expected as light-flavor production at low $p_\mathrm{T}$ does not scale with $N_\mathrm{coll}$ 
and is also observed when comparing to cocktail ratios that include scaling of light flavor in p$+$Pb.

\begin{figure}[t]
    \centering
    \includegraphics[width=.38\textwidth]{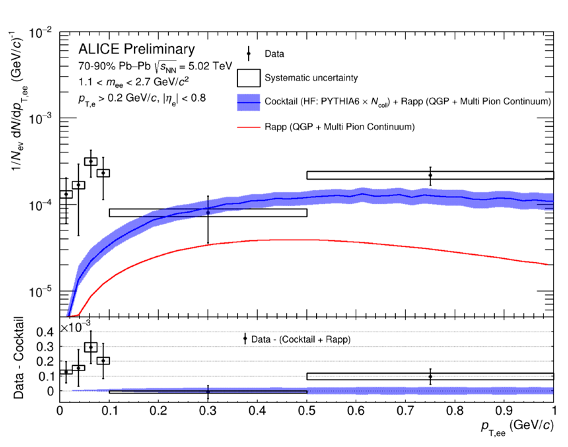}
    \includegraphics[width=.42\textwidth]{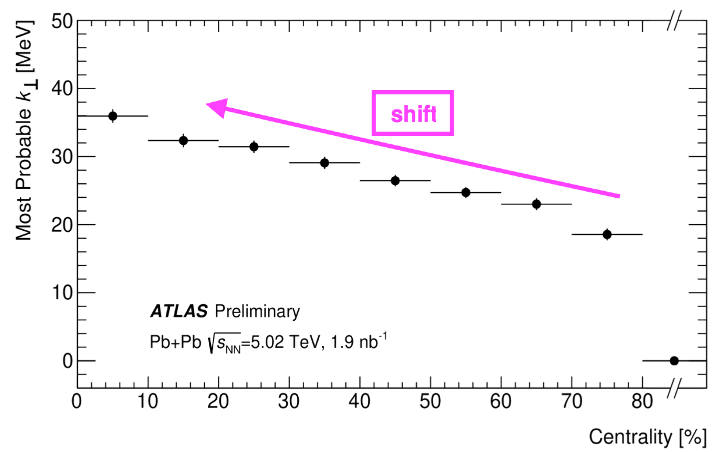}
    \caption{Low-$p_\mathrm{T}$ dilepton measurements in peripheral Pb$+$Pb collisions at $\sqrt{s_\mathrm{NN}}=5.02$~TeV. Left panel: ALICE low-$p_\mathrm{T}$
    dielectron measurements \cite{qm2019:ALICE:Scheid}. Right panel: centrality dependence of $k_\perp$
    from dimuon measurement from ATLAS \cite{qm2019:ATLAS:Palni}.
    }
    \label{fig:lowptdileptons}
\end{figure}

Coherent $\gamma$-N and $\gamma-\gamma$ interactions are conventionally studied in UPC interactions. Recently, the STAR
and ATLAS experiments have published observations of low-$p_\mathrm{T}$ dilepton excess in hadronic heavy-ion collisions 
\cite{Adam:2018tdm,Aaboud:2018eph}.
In the left panel of Fig.~\ref{fig:lowptdileptons} new measurements in Pb$+$Pb collisions at 5.02~TeV from the ALICE 
collaboration show a $3\sigma$ low-$p_\mathrm{T}$ excess in the intermediate mass range for the 70-90\% centrality class. Meanwhile, 
both STAR and ATLAS have further expanded their measurements by including low-$p_\mathrm{T}$ dimuons, and combining the 5.02~TeV data 
sets, respectively.  Based on the combined 2014 and 2018 data sets, the ATLAS collaboration showed a distinct centrality
dependence of $\langle k_\perp\rangle$ in Pb$+$Pb collisions at $\sqrt{s_\mathrm{NN}}=5.02$~TeV, as can be seen in the right
panel of Fig.~\ref{fig:lowptdileptons} \cite{qm2019:ATLAS:Palni}.

\begin{figure}[h]
    \centering
    \includegraphics[width=.3\textwidth]{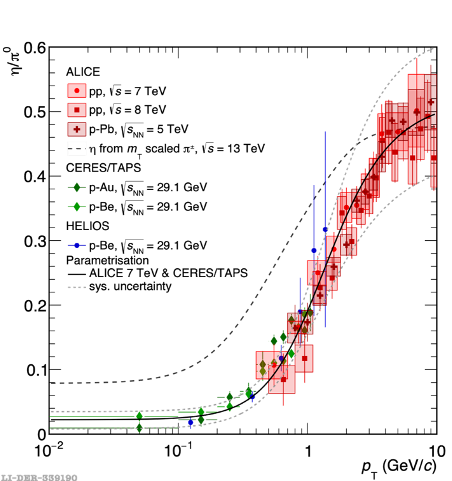}
    \includegraphics[width=.34\textwidth]{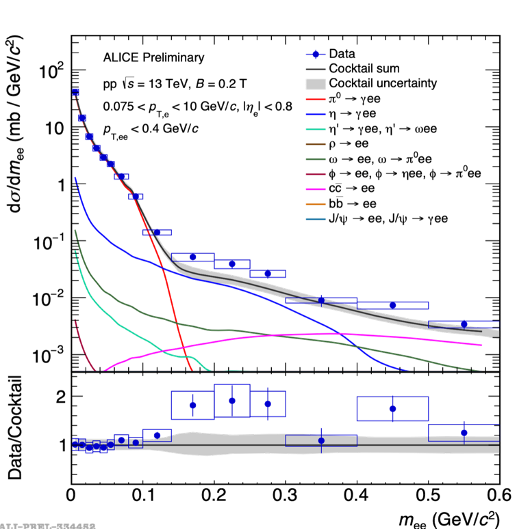}
    \includegraphics[width=.34\textwidth]{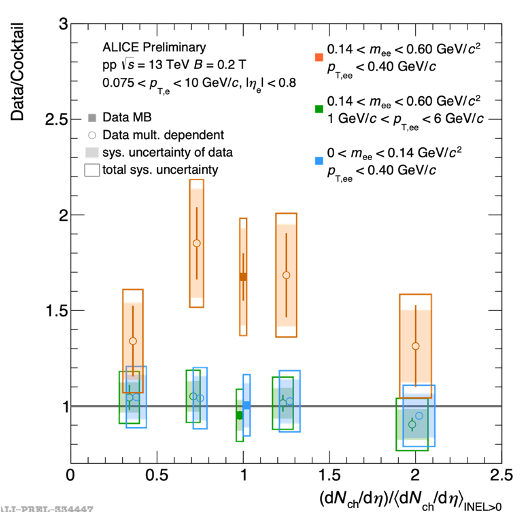}
    \caption{Soft dielectron production in p$+$p collisions at $\sqrt{s}=13$~TeV measured by 
    ALICE \cite{qm2019:ALICE:Scheid}. Left panel: new parametrization of the $\eta/\pi$ ratio. Middle panel: 
    low dielectron invariant-mass spectrum in p$+$p collisions at $\sqrt{s}=13$~TeV. Right panel: data-over-cocktail 
    ratio for low-$p_\mathrm{T}$ $\eta$ mesons compared to $\pi^0$ mesons in the same low-$p_\mathrm{T}$ range, 
    $\eta$ mesons in a  higher $p_\mathrm{T}$ range.}
    \label{fig:softdielectrons}
\end{figure}

At the 2018 Quark Matter conference, the ALICE collaboration reported on its potential for studying in p$+$p
collisions at $\sqrt{s}=13$~TeV a soft dielectron enhancement \cite{Bailhache:2019wyf} that was first reported
at the ISR by the Axial Field Spectrometer collaboration for p$+$p at $\sqrt{s}=63$~GeV \cite{Hedberg:1987yq}. At
the time of the previous conference, large uncertainties on the contribution of the $\eta$ meson to the hadronic cocktail and
limited statistics did not allow for a quantitative conclusion. At this conference, the collaboration reported on 
its findings from a special run in which the field in its solenoid magnet was lowered to $B=0.2$~T. This allows 
its low-$p_\mathrm{T}$ reach for electrons to drop to 75~MeV/$c$. Additionally, ALICE presented a reevaluation of the 
$\eta$ contribution as is shown in the left panel of Fig.~\ref{fig:softdielectrons}. Combined with the new low B-field run,
these improvements now show a significant enhancement over the cocktail for $p_{T,\mathrm{ee}} < 0.4$~GeV/$c$ in the $\eta$
mass range as can be seen in the middle panel. Interestingly, and shown in the right panel of Fig.~\ref{fig:softdielectrons},
a comparison of the data for $\pi^0$s in a similar momentum range and $\eta$ mesons at higher $p_\mathrm{T}$ is consistent with the
hadronic cocktail calculations. The physical mechanism for this enhancement is not yet understood.

\section{The Near-Future Experimental Landscape}
\label{sect:future}
Measurements of electromagnetic and weak probes have posed particular challenges to experiments in the past and 
present in terms of {\em e.g.} material budgets and event statistics. At this conference several experiments, located
at various facilities, presented updates on the current state of their general upgrade plans and future designs. Many
of the planned upgrades and designs specifically relate to improvements in the measurements that are relevant to the topic
of this paper. At the same time, several collaborations showed very encouraging glimpses into what to expect from
very recently collected data sets. Starting at the lower center-of-mass energies, the HADES collaboration showed its first
raw dielectron spectra, collected in 2019, of Ag$+$Ag collisions at 2.42~GeV and 2.55~GeV with 1.3 and 14 billion events,
respectively \cite{qm2019:HADES:Harabaz}. The STAR collaboration embarked on the second phase of its BES program where it 
successfully collected\footnote{at the time of this writing} large data sets of Au$+$Au collisions at
$\sqrt{s_\mathrm{NN}}=19.6$1, 14.6, and 11.5~GeV. These, and the proposed 9.2 and 7.7~GeV data sets will allow the 
collaboration to complete its beam energy scan of the low $M_\mathrm{ee}$ excess yields in a range where the total baryon 
density is expected to increase as beam energies are lowered \cite{Adam:2018qev}. The BES Phase-2 data sets will provide
sufficient statistics in the intermediate mass range to allow a simultaneous extraction of the medium temperature.
\begin{figure}[h!]
    \centering
    \includegraphics[width=0.43\textwidth]{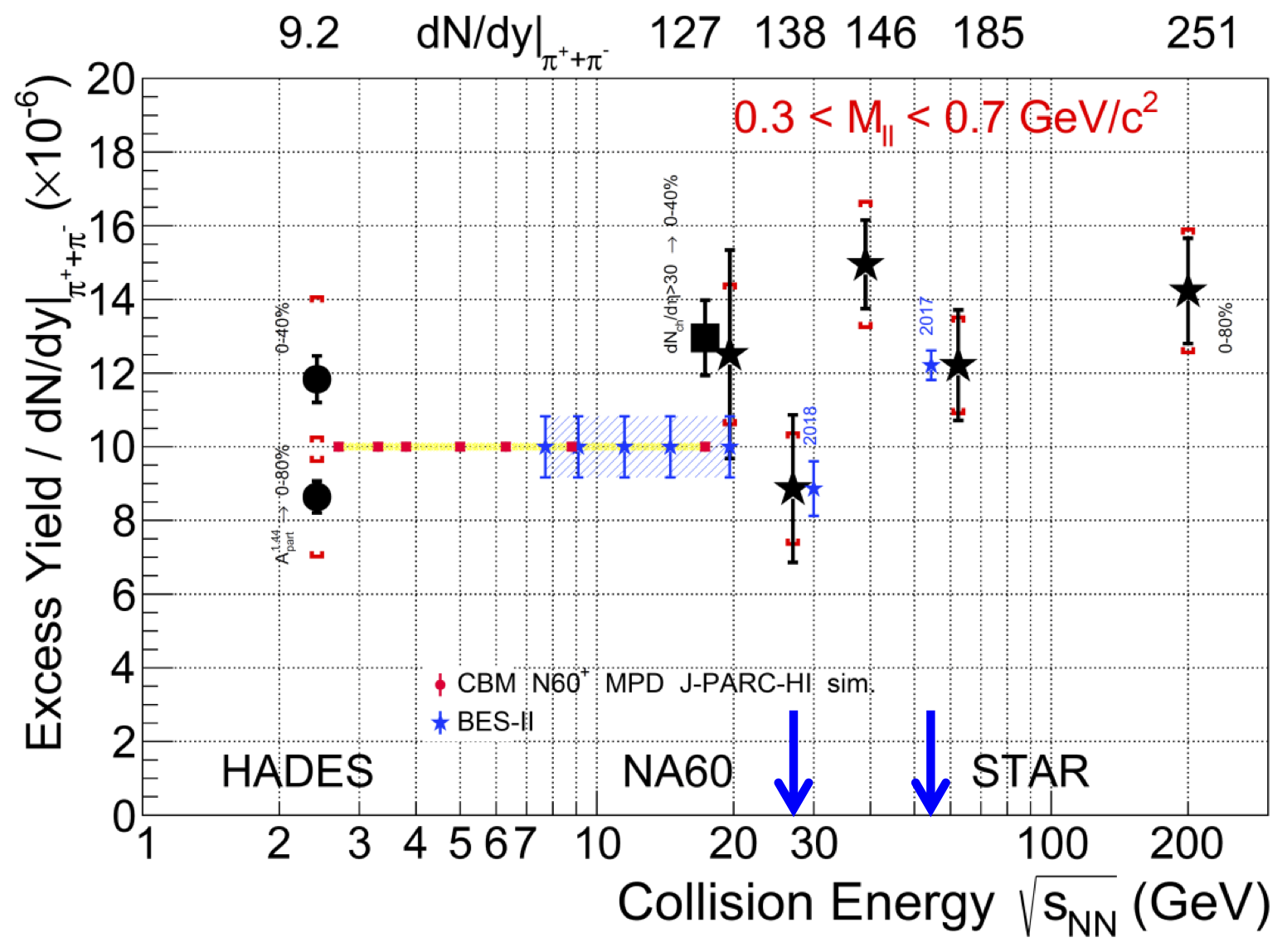}
    \includegraphics[width=0.56\textwidth]{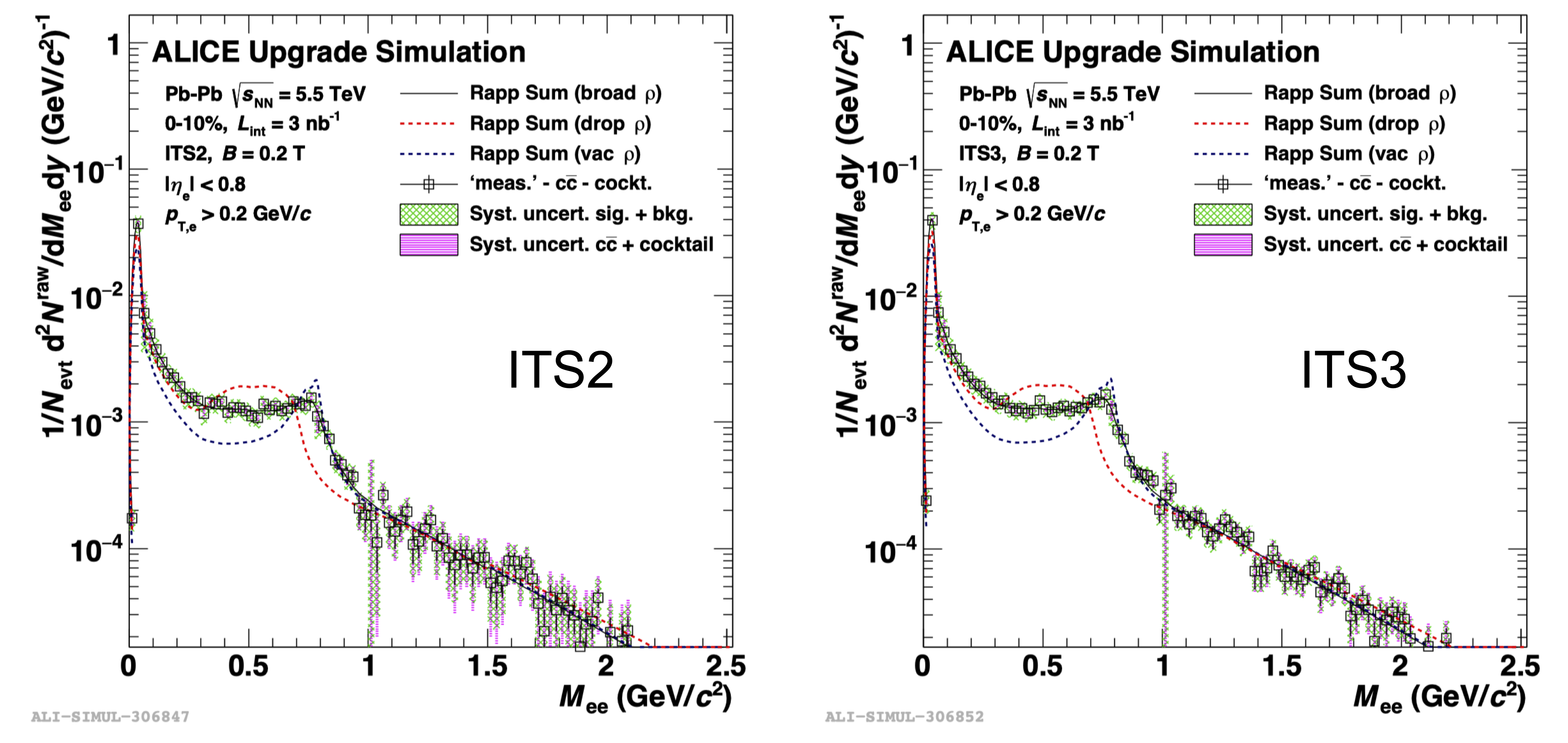}
    \caption{Left panel: excess yields scaled by $dN/dy|_{\pi^\pm}$ from SIS to RHIC energies \cite{qm2019:STAR:Seck}.
              Right and middle panels: ALICE ITS2 and ITS3 upgrade simulations of dielectron measurements in 
              Pb$+$Pb \cite{qm2019:ALICE:Reidt}.}
    \label{fig:future}
\end{figure}{}

In the left panel of Fig.~\ref{fig:future}, normalized excess yields of HADES, NA60, and STAR
\cite{Adamczewski-Musch:2019byl,qm2019:STAR:Seck,Adam:2018qev,Specht:2010xu,Adamczyk:2015mmx} are combined with the
energies and projected precisions of four proposed, future detectors. The MPD detector at NICA will install its ECAL 
detector starting 2020 which will complement its proposed program to measure electromagnetic probes in addition to the 
conversion-based methods \cite{qm2019:MPD:Kisiel}. The proposed E16 pilot run at J-PARC \cite{qm2019:JPARC:Sako} aims to 
use its dielectron spectrometer to study vector-meson modifications in p$+$A collisions at very high rates. The CBM experiment
at FAIR is configurable for either dimuon of dielectron measurements and is expected slated for operations in 
2025 \cite{qm2019:CBM:Klochkov}. At the SPS, the NA60+ collaboration proposes a detector that will allow for high-precision,
high statistics measurements of the full dimuon spectrum, which will include the opportunity to measure the effects of 
chiral $\rho-a_1$ mixing.

At higher energies, the LHC experiments presented the status of their upgrades for the upcoming Run-3 and plans for Run-4.
Proposed LHCb upgrades will enable Drell-Yan measurements from low-mass dimuon at forward rapidities which can probe the 
gluon nPDF at small $x$ \cite{qm2019:LHCb:Nezza}. Of particular interest to the dilepton measurements, the ALICE collaboration
presented its expectations for Run 3. The new ITS will significantly improve the tracking resolution, while the TPC readout 
upgrade would increase the data rate by two orders of magnitude. Simulations, shown in the middle and right panels of 
Fig.~\ref{fig:future}, demonstrate the significant improvements that these upgrades will bring to the Run-3 and Run-4
dielectron measurements, respectively. 

In conclusion, electromagnetic and weak probes are the most versatile probes available in heavy-ion collisions with precision 
data collected over almost four orders of magnitude in collision energies. Many upgrades and new detector proposals make the
future of weak and electromagnetic probes truly look very bright.

\section*{Acknowledgements}
The author would like to thank the organizers of the Quark Matter 2019 conference for the invitation to present this overview. This work
is in part supported by the U.S. Department of Energy under grant No.~DE-FG02-10ER41666.

\bibliographystyle{elsarticle-num}
\bibliography{references.bib}

\begin{thebibliography}{10}
\expandafter\ifx\csname url\endcsname\relax
  \def\url#1{\texttt{#1}}\fi
\expandafter\ifx\csname urlprefix\endcsname\relax\def\urlprefix{URL }\fi
\expandafter\ifx\csname href\endcsname\relax
  \def\href#1#2{#2} \def\path#1{#1}\fi

\bibitem{Armesto:2006ph}
N.~Armesto, {Nuclear shadowing}, J. Phys. G32 (2006) R367--R394.

\bibitem{qm2019:LHCb:Li}
H.~Li, Z production in {pPb} collisions and charmonium production in {Pb-Pb}
  ultra-peripheral collisions at {LHCb}, in: these proceedings
  \cite{qm2019proceedings}.

\bibitem{qm2019:ALICE:Weber}
M.~Weber, Highlights from {ALICE}, in: these proceedings
  \cite{qm2019proceedings}.

\bibitem{Sirunyan:2019dox}
A.~M. Sirunyan, et~al., {Observation of nuclear modifications in W$^\pm$ boson
  production in pPb collisions at $\sqrt{s_\mathrm{NN}} =$ 8.16 TeV}, Phys.
  Lett. B800 (2020) 135048.

\bibitem{Aad:2019sfe}
G.~Aad, et~al., {Measurement of $W^\pm $ boson production in Pb+Pb collisions
  at $\sqrt{s_{\mathrm{NN}}} $= 5.02~TeV with the ATLAS detector}, Eur. Phys.
  J. C79~(11) (2019) 935.

\bibitem{David:2019wpt}
G.~David, {Direct real photons in relativistic heavy ion collisions }\href
  {http://arxiv.org/abs/1907.08893} {\path{arXiv:1907.08893}}.

\bibitem{qm2019:PHENIX:Fan}
W.~Fan, Thermal photon production in {Au+Au} collisions {(PHENIX)}, in: these
  proceedings \cite{qm2019proceedings}.

\bibitem{Adare:2014fwh}
A.~Adare, et~al., {Centrality dependence of low-momentum direct-photon
  production in Au$+$Au collisions at $\sqrt{s_{_{NN}}}=200$ GeV}, Phys. Rev.
  C91~(6) (2015) 064904.

\bibitem{Adare:2008ab}
A.~Adare, et~al., {Enhanced production of direct photons in Au+Au collisions at
  $\sqrt{s_{NN}}=200$ GeV and implications for the initial temperature}, Phys.
  Rev. Lett. 104 (2010) 132301.

\bibitem{Adare:2011zr}
A.~Adare, et~al., {Observation of direct-photon collective flow in
  $\sqrt{s_{NN}}=200$ GeV Au+Au collisions}, Phys. Rev. Lett. 109 (2012)
  122302.

\bibitem{Adare:2018wgc}
A.~Adare, et~al., {Beam Energy and Centrality Dependence of Direct-Photon
  Emission from Ultrarelativistic Heavy-Ion Collisions}, Phys. Rev. Lett. 123
  (2019) 022301.

\bibitem{STAR:2016use}
L.~Adamczyk, et~al., Direct virtual photon production in {Au}+{Au} collisions
  at $\sqrt{s_\mathrm{NN}}$ = 200 {GeV}, Phys. Lett. B 770 (2017) 451.

\bibitem{Adamczewski-Musch:2019byl}
{The HADES Collaboration}, Probing dense baryon-rich matter with virtual
  photons, Nature Physics 15 (2019) 1040.

\bibitem{Arnaldi:2006jq}
R.~Arnaldi, et~al., {First measurement of the rho spectral function in
  high-energy nuclear collisions}, Phys. Rev. Lett. 96 (2006) 162302.

\bibitem{qm2019:HADES:Harabaz}
S.~Harabasz, Characterizing baryon dominated matter with {HADES} measurements,
  in: these proceedings \cite{qm2019proceedings}.

\bibitem{Chatterjee:2007xk}
R.~Chatterjee, D.~K. Srivastava, U.~W. Heinz, C.~Gale, {Elliptic flow of
  thermal dileptons in relativistic nuclear collisions}, Phys. Rev. C75 (2007)
  054909.

\bibitem{Adamczyk:2014lpa}
L.~Adamczyk, et~al., Dielectron azimuthal anisotropy at mid-rapidity in {Au} +
  {Au} collisions at $\sqrt{s_\mathrm{NN}} = 200$ {GeV}, Phys. Rev. C 90 (2014)
  064904.

\bibitem{qm2019:STAR:Seck}
F.~Seck, Measurements of dielectron production in {Au+Au} collisions at
  $\sqrt{s_{NN}}=27$, 54.4 and 200~{GeV} with the {STAR} experiment, in: these
  proceedings \cite{qm2019proceedings}.

\bibitem{qm2019:ALICE:Scheid}
S.~Scheid, Low-mass dielectron measurements in pp, {p-Pb} and {Pb-Pb}
  collisions with {ALICE} at the {LHC}, in: these proceedings
  \cite{qm2019proceedings}.

\bibitem{Adam:2018qev}
J.~Adam, et~al., {Measurements of Dielectron Production in Au$+$Au Collisions
  at $\sqrt{s_\mathrm{NN}}$= 27, 39, and 62.4 GeV from the STAR Experiment
  }\href {http://arxiv.org/abs/1810.10159} {\path{arXiv:1810.10159}}.

\bibitem{qm2019:ATLAS:Palni}
P.~Palni, Measurements of dileptons and photon pairs from two-photon scattering
  in ultra-peripheral and hadronic {Pb+Pb} collisions with {ATLAS} detector,
  in: these proceedings \cite{qm2019proceedings}.

\bibitem{Adam:2018tdm}
J.~Adam, et~al., Low-$p_{T}$ $e^+e^-$ pair production in {Au}$+${Au} collisions
  at $\sqrt{s_{NN}}$ = 200 {GeV} and {U}$+${U} collisions at $\sqrt{s_{NN}}$ =
  193 {GeV} at {STAR}, Phys. Rev. Lett. 121 (2018) 132301.

\bibitem{Aaboud:2018eph}
M.~Aaboud, et~al., {Observation of centrality-dependent acoplanarity for muon
  pairs produced via two-photon scattering in Pb+Pb collisions at
  $\sqrt{s_{\mathrm{NN}}}=5.02$ TeV with the ATLAS detector}, Phys. Rev. Lett.
  121 (2018) 212301.

\bibitem{Bailhache:2019wyf}
R.~Bailhache, {Dielectron measurements in pp and Pb–Pb collisions with ALICE
  at the LHC}, Nucl. Phys. A982 (2019) 779.

\bibitem{Hedberg:1987yq}
V.~Hedberg, {Production of positrons with low transverse momentum and low mass
  electron - positron pairs in proton proton collisions at a center-of-mass
  energy of 63-GeV}, Ph.D. thesis, Lund U. (1987).

\bibitem{qm2019:ALICE:Reidt}
F.~Reidt, Upgrading the {Inner Tracking System} and the {Time Projection
  Chamber} of {ALICE}, in: these proceedings \cite{qm2019proceedings}.

\bibitem{Specht:2010xu}
H.~J. Specht, {Thermal Dileptons from Hot and Dense Strongly Interacting
  Matter}, AIP Conf. Proc. 1322 (2010) 1.

\bibitem{Adamczyk:2015mmx}
L.~Adamczyk, et~al., Energy dependence of acceptance-corrected dielectron
  excess mass spectrum at mid-rapidity in {Au}$+${Au} collisions at
  $\sqrt{s_{NN}} =$ 19.6 and 200 {GeV}, Phys. Lett. B 750 (2015) 64.

\bibitem{qm2019:MPD:Kisiel}
A.~Kisiel, The physics performance of the {MPD} detector at {JINR}, in: these
  proceedings \cite{qm2019proceedings}.

\bibitem{qm2019:JPARC:Sako}
H.~Sako, Exploring dense baryonic matter and multi-strangeness at {J-PARC
  Heavy-Ion Project}, in: these proceedings \cite{qm2019proceedings}.

\bibitem{qm2019:CBM:Klochkov}
V.~Klochkov, The {Compressed Baryonic Matter} {(CBM)} experiment at {FAIR}, in:
  these proceedings \cite{qm2019proceedings}.

\bibitem{qm2019:LHCb:Nezza}
P.~D. Nezza, {LHC} {Run 3} and {Run 4} prospects for heavy-ion physics with
  {LHCb}, in: these proceedings \cite{qm2019proceedings}.

\bibitem{qm2019proceedings}
QM2019 Proceedings.

\end{thebibliography}

\end{document}